\documentstyle[12pt]{article}
\topmargin--0cm
\oddsidemargin--1mm
\begin{document}
\newcommand{\pl}{\partial}
\newcommand{\mb}[1]{\mbox{\boldmath${\bf #1}$}}

\begin{center}
{\LARGE Dynamical Abelian Projection of Gluodynamics}
\footnote{A talk given at QCD96, Montpellier, France (July, 1996)}

\vskip 0.5cm
{Sergei V. SHABANOV}\footnote{On leave from {\em Laboratory of
Theoretical Physics, JINR, Dubna, Russia}}

\vskip 0.5cm
{\em Department of Theoretical Physics, University of Valencia}\\
{\em Dr. Moliner 50, Burjassot (Valencia), E-46100, Spain}
\end{center}

\begin{abstract}
Assuming the monopole dominance, that has been proved
in the lattice gluodynamics, to hold in the continuum
limit, we develop an effective scalar field theory
for QCD at large distances to describe confinement.
The approach is based on a gauge (or projection)
independent formulation of the monopole dominance and
manifestly Lorentz invariant.
\end{abstract}

\subsection*{1. The monopole dominance}

Numerical simulations \cite{stack}
show that there exist configurations $\bar{A}_\mu$ of gauge fields
which dominate in the path integral for the Wilson loop expectation
value, i.e. they give a main contribution to the QCD string tension
and, therefore, are the most relevant ones for the confinement in QCD.
In fact, if one calculates the Wilson
loop average over a subset formed by these specific configurations, the
difference between the string tension extracted from such average
and the full QCD string tension (extracted from the Wilson loop average
over all possible gauge field configurations) appears to be about
eight per cent.

When taken in a specific gauge $\chi(A) =0$ that breaks
the gauge group $G$ to its maximal Abelian subgroup $G_H$
(e.g. SU(3) to U(1)$\times$U(1)),
the dominant  configurations are Dirac magnetic monopoles with
respect to the unbroken Abelian gauge group \cite{thooft}. For
this reason, the phenomenon is called the monopole dominance.
It is should be noted that not in every {\it Abelian projection}
$\chi$, the dominant configurations are monopoles \cite{polikarpov}.
If the dominant configurations are selected
as those that turn into magnetic monopoles
when lifted by a gauge transformation on the surface $\chi(A) =0$,
the monopole dominance may or may not occur, depending on the choice
of $\chi$. This has been indeed observed in the lattice simulations
\cite{pisa}. In fact, the monopole dominance has been found, up to now,
only in the so called maximal Abelian projection. In the continuum limit,
this gauge condition has the form
\begin{equation}
D_\mu (A^H)A^{off}_{\mu}= \pl_\mu A_\mu^{off} + ig[A^H_\mu ,
A^{off}_{\mu}] =0\ ,
\label{00}
\end{equation}
where
$A_\mu =A^H_\mu +A^{off}_\mu$ and $A^{H,off}_{\mu}$ are Abelian (diagonal)
and off-diagonal components of $A_\mu$, respectively. So, not every
Abelian projection can be used to select configurations relevant for the
QCD confinement.

\subsection*{2. Monopoles and Abelian projections}

In what follows, the gauge
group is always assumed to be SU(2) and, hence, $G_H=U(1)$. It allows one
to avoid some unnecessary technicalities. A generalization does
not meet any difficulty.
To describe the dominant configurations via an Abelian projection, one has
to choose a gauge condition $\chi$ that breaks SU(2) to $U(1)$. Given
$A_\mu $, one should find a gauge group element
$\Omega _\chi =\Omega _\chi (A)$ such that the gauge transformed
configuration
\begin{equation}
A^{\Omega _\chi}_\mu =\Omega _\chi A_\mu \Omega ^\dagger_\chi
+i/g\ \Omega _\chi \pl _\mu \Omega ^\dagger_\chi
\label{0}
\end{equation}
lies on the surface $\chi =0$,
i.e. $\Omega _\chi $ satisfies the equation
\begin{equation}
\chi (A^{\Omega _\chi })=0\ .
\label{1}
\end{equation}
Suppose one can solve (\ref{1}) for a generic $A_\mu$.
The abelian (Maxwell) field $C^\chi _\mu $
associated with the unbroken $U(1)$ group is extracted as follows
\begin{eqnarray}
A^{\Omega _\chi }_\mu &= & W^\chi _\mu +\tau _3C^\chi _\mu /2 \ ,
\label{2} \\
C^\chi _\mu &= & {\rm tr}\ \tau _3(\Omega _\chi A_\mu \Omega
^\dagger_\chi + i/g\ \Omega _\chi \pl _\mu \Omega ^\dagger_\chi)\ ,
\label{3}
\end{eqnarray}
where ${\rm tr} \tau _3W^\chi _\mu \equiv 0$,
and $\tau_a$ are the Pauli matrices, $tr \tau_b \tau_a =2\delta_{ab}$.
 Following the lattice
procedure, one picks up a space point $\mb{x}$ (time is fixed) and
surrounds it by a sphere $\Sigma _x$ centered at $\mb{x}$. On the
sphere, one takes an infinitesimal closed contour $L(\mb{x}_s)$ centered
at $\mb{x}_s\in \Sigma _x$ and calculates a flux of the magnetic
field $\mb{B}^\chi={\rm curl}\ \mb{C}^\chi$ through
$L(\mb{x}_s)$ in the limit when $L$ shrinks to $\mb{x}_s$. If $C^\chi _\mu
$ is regular everywhere on $\Sigma_x$, the flux always vanishes.
If the flux happens to be non-zero for some
$\mb{x}_s\in \Sigma _x$, then there are Dirac strings passing through
$\mb{x}_s$. Repeating this procedure for all $\mb{x}$, one can locate a
net of Dirac strings and, hence,
determine a distribution of monopoles associated
with the configuration $A_\mu$ and the gauge fixing $\chi$.

The procedure should be applied to all configurations $A_\mu \in
[A]$ to obtain all possible monopole configurations in the Abelian
projection $\chi$. Configurations in $[A]$ that give no monopole after
the projection are assumed to be irrelevant and can be thrown from the
sum over configurations in the Wilson loop average.

Suppose there is a monopole at $\mb{x}$. On the sphere $\Sigma _x$, the
first term in (\ref{3}) is regular so is the magnetic field associated
with it. Therefore it does not give a finite contribution to the flux
through an infinitesimal surface cut out from $\Sigma_{x}$ by the contour
$L(\mb{x}_s)$, whereas, according to the Stocks theorem, the second term in
(\ref{3}) may give a finite contribution to the flux because it can contain
a total derivative of multi-valued angular functions which parametrize the
group element $\Omega _\chi $. That is, the net of Dirac strings as well as
position of monopoles are completely specified by all $\Omega _\chi $.
Since the total magnetic flux through $\Sigma _x$ must be zero (the flux
of the magnetic Coulomb field of a monopole at $\mb{x}$ is equal to the
flux carried by the Dirac string of the monopole), we obtain for the
magnetic charge at $\mb{x}$
\begin{eqnarray}
q_\chi (\mb{x})&=&\frac{i}{4\pi}\oint_{\Sigma _x}\! d\sigma _j
\varepsilon _{jkn}{\rm tr} (e_\chi [\pl _ke_\chi , \pl _ne_\chi ])\ ,
\label{4}\\
e_\chi &=&\Omega ^\dagger_\chi \tau _3\Omega _\chi\ ,
\ \ \ \ {\rm tr}\ e^2_\chi =2\ ,
\label{4a}
\end{eqnarray}
where
the radius of $\Sigma _x$ tends to zero. The integer (\ref{4})
classifies maps $\Sigma _x \rightarrow SU(2)/U(1)\sim S^2$ carried
out by the field $e_\chi $.
It determines how many times the sphere $\Sigma _x$ is
wrapped around the sphere $SU(2)/U(1)\sim S^2$.

The lattice simulations show \cite{suzuki} that the QCD string tension
does not depend on the off-diagonal element $W^\chi _\mu $ (the Abelian
dominance); they are set to be zero $W^\chi _\mu =0$ after the projection
and are not accounted for in the Wilson loop average over monopoles.
Photon configurations of the Maxwell  field $C^\chi _\mu $
(monopole-free configurations) are also irrelevant for the string tension
(the monopole dominance \cite{stack}), i.e. only the last monopole term in
(\ref{3}), denoted below as $\bar{C}^\chi _\mu$, is important for the
confinement. Thus, the monopole dominance implies that the dominant
configurations can be parametrized by the set $[\Omega _\chi ]$ of
$\Omega _\chi$'s with $q_\chi \neq 0$ at least at one point $\mb{x}$.
A parametrization of the set $[\Omega_\chi]$ is technically difficult to
find. It implies solving equation (\ref{1}) for a generic $A_\mu$
which seems hardly possible
in the maximal Abelian projection (\ref{00})
where the monopole dominance is shown
to occur. An alternative approach has to be developed in the continuum
theory.

It should be noted that in the continuum limit, one cannot
simply put $W^\chi _\mu =0$ and count only contributions
of pure monopole configurations in the Wilson loop average
because Dirac monopoles are pointlike objects and, hence, have infinite
magnetic field energy.
Since the string tension is not sensitive to a specific
form of $W_\mu^\chi$, we choose $W^\chi _\mu =\bar{W}^\chi _\mu
(\bar{C}^\chi)$ to provide a finite size to the Dirac monopoles
$\bar{C}^\chi _\mu =-i/g\ {\rm tr} (e_\chi \Omega^\dagger_\chi \pl _\mu \Omega
_\chi )$. The core functions $\bar{W}^\chi _\mu$ depend only on the
monopole distributions (\ref{4}).
The configurations $\bar{W}^\chi _\mu +\tau_3\bar{C}_\mu /2$ are known
as the Wu-Yang monopoles with a core and their explicit construction is
given in \cite{kogut}. The full color {\em magnetic} energy $\int d^3x{\rm
tr}F^2_{ij}/4$ is finite for these configurations. Therefore the action
remains also finite at finite temperature. Note that in the lattice QCD,
the lattice spacing plays the role of the monopole energy regularization.

As has been aforementioned, the Abelian projection is used only
to select the dominant configurations. It does not offer any explanation
why these configurations are dominant. In this regard, the monopole
dominance is nothing but an "experimental" fact discovered numerically.
Nevertheless, this fact can be exploited
to classify the dominant configurations.
To do so, we lift back the regularized monopole configurations $\bar{W}^\chi
_\mu (\bar{C}^\chi )+\bar{C}^\chi _\mu $ from the surface $\chi =0$ to the
space of all configurations $[A]$ by a gauge transformation with $\Omega
=\Omega ^\dagger _\chi$. We denote the image of the lift $[\ ^\chi
\bar{A}]\subset [A]$. By construction, any gauge potential $\ ^\chi
\bar{A}_\mu $ from $[ ^\chi \bar{A}]$ becomes a monopole with a
core when projected onto the gauge fixing surface $\chi =0$, meaning that
\begin{equation}
 ^\chi \bar{A}^{\Omega _\chi}_\mu =\bar{W}^\chi _\mu +\bar{C}^\chi _\mu\ .
\label{4b}
\end{equation}

The subset $[\ ^\chi \bar{A}]$ is the projection-dependent (or
gauge-dependent) and defined up to {\it regular}
gauge transformations of its
elements. The latter arbitrariness is associated with the freedom to
redefine $\Omega_\chi$ by a shift on a regular group element,
$\Omega_\chi\rightarrow\Omega_\chi\Omega_0$,
such that the distribution of monopoles (\ref{4})
remains untouched.

Consider a {\em gauge invariant} (projection independent) subset
\begin{equation}
[\bar{A}] =\cup_\chi [ ^\chi \bar{A}]  \subset [A]\ ,
\label{5}
\end{equation}
which is the union of the subsets $[ ^\chi \bar{A}] $
found in all possible
Abelian projections. By definition
a gauge potential $\bar{A}_\mu$ belongs to $[\bar{A}]$
if there exists an Abelian projection $\chi$ such that
$\bar{A}^{\Omega_\chi}_\mu$ is a magnetic monopole with a core and
$\Omega_\chi$ satisfies (\ref{1}). The projection independent set
$[\bar{A}]$ is formed by configurations that dominate at large distances
and are responsible for generating the QCD string between two static
sources. This conjecture is supported by lattice simulations. The
maximal Abelian projection seems to catch a major part of $[\bar{A}]$,
while some other projections do not. The latter explains the dependence
of the monopole dominance on the projection recipe.

Our next problem is to find an appropriate parametrization of $[\bar{A}]$
and develop an effective field theory for dynamics in it.

\subsection*{\bf 3. Universality of the dynamical Abelian projection}

In \cite{sh}
the dynamical Abelian projection has been proposed. This projection
does not rely on any specific gauge condition $\chi(A)=0$ to break
the gauge group to its maximal Abelian subgroup. The idea was to
insert the identity
\begin{equation}
\sqrt{\det (-D_\mu^2)}\!\int {\cal D}\phi \exp i
\!\! \int d^4x {\rm tr}
(D_\mu \phi)^2 =1,
\label{6}
\end{equation}
where a real scalar field $\phi$ realizes the adjoint representation
of the gauge group, into the integral over gauge fields.
Now the gauge fields are coupled to an auxiliary scalar field in
a gauge invariant way via the covariant derivative. It can be exploited
to achieve an Abelian
projection by imposing the following gauge condition on $\phi$: The
off-diagonal components of $\phi$ are set to be zero. Positions of
Dirac monopoles are determined by zeros of a gauge invariant polynom
of $\phi$ that coincides with the Faddeev-Popov determinant in the
unitary gauge imposed on $\phi$ \cite{sh}.

To perform the dynamical Abelian projection for SU(2), one should
solve (\ref{1}) which assumes the form ${\rm tr}(\tau_3 \Omega_\phi
\phi\Omega_\phi^\dagger) =0$, i.e., given $\phi$, we look for $\Omega_\phi
\in$ SU(2) whose adjoint action on $\phi$ brings it to a diagonal
form. The group element $\Omega_\phi$ is ill-defined at spacetime
points where $\phi(x) =0$. The latter condition implies three
equations on four spacetime coordinates. Their solutions therefore
determine world lines $x^\mu = x^\mu(\tau)$ which are shown to be
world lines of magnetic monopoles \cite{sh}.
To see it, we remark first that
at fixed time all singularities of a generic $\Omega_\phi$ form a
set of isolated points in space. Gauge potentials in the dynamical
Abelian projection assume the form (\ref{0}) with $\Omega_\chi
=\Omega_\phi$. Let $\phi=0$ at $\mb{x}$. Applying the monopole
location procedure of section 2 to the Maxwell potential $C^\phi_\mu ={\rm
tr}(\tau_3 A^{\Omega_\phi}_\mu)$, we find that the magnetic charge of
the monopole at $\mb{x}$ is equal to
$q_\phi(\mb{x})$ given by (\ref{4}) where
$e_\chi =e_\phi= \Omega_\phi^\dagger\tau_3\Omega_\phi$.

According to (\ref{6}) the scalar field fluctuates and therefore
the set $[\Omega_\phi]$ of all $\Omega_\phi$'s covers
{\em all possible} monopole configurations. Note that any distribution
of monopoles $q_\phi(x)$ is determined by some $\Omega_\phi$ in
(\ref{4}). So, we conclude
\begin{equation}
[ ^\chi\bar{A}]\sim [\Omega_\chi] \subseteq [\Omega_\phi]
\label{7}
\end{equation}
for any Abelian projection $\chi$. For a fixed distribution $q_\phi(x)$,
consider a gauge potential $ ^\phi \bar{A}_\mu$ such that
$ ^\phi\bar{A}^{\Omega_\phi}_\mu = \bar{W}_\mu^\phi (\bar{C}) +
\tau_3\bar{C}_\mu^\phi/2$, where $\bar{C}_\mu^\phi= -i/g\ {\rm tr}
(e_\phi\Omega_\phi^\dagger\pl_\mu\Omega_\phi)$ is the Dirac monopole potential
for given $q_\phi(x)$ and $\bar{W}_\mu^\phi$ are associated core
functions. Let $[ ^\phi\bar{A}]\subset [A]$ be a set of configurations
$ ^\phi\bar{A}_\mu$ for all possible monopole distributions $q_\phi(x)$.
Clearly, $[ ^\phi\bar{A}]$ is isomorphic to $[\Omega_\phi]$, and from
(\ref{7}) follows that $[ ^\phi\bar{A}]$ is larger than
$[ ^\chi\bar{A}]$  for
any $\chi$. Thus, the set $[ ^\phi\bar{A}]$
covers the gauge invariant set (\ref{5}):
\begin{equation}
[\bar{A}] \subseteq [^\phi\bar{A}]\subset [A]\ .
\label{8}
\end{equation}

Thereby, the dominant configurations $[\bar{A}] $ can be parametrized by
the $\Omega_\phi$'s.

\subsection*{\bf 4. The effective scalar field theory}

To develop an effective
theory for dynamics in the dominant sector $[\bar{A}]$,
we substitute the identity (\ref{6}) in the path integral
for the Wilson loop average and perform the change of variables in it
\begin{equation}
A_\mu =\ ^\phi\bar{A}_\mu(\phi) + a_\mu\ ,
\label{9}
\end{equation}
where $ ^\phi\bar{A}_\mu\in [^\phi\bar{A}]$ has been described above.
Since (\ref{9}) is just a shift of a generic gauge field
con\-fi\-gu\-ration
on a fixed con\-fi\-gu\-ra\-tion pa\-ra\-met\-ri\-zed by
an in\-de\-pen\-dent integration
variable $\phi$, the mea\-su\-re as\-sumes
the form ${\cal D}A_\mu{\cal D}\phi =
{\cal D}a_\mu{\cal D}\phi$.

Assuming that the monopole dominance holds in the continuum limit, we
can integrate out the small fluctuations $a_\mu$ around the dominant
configurations $ ^\phi\bar{A}(\phi)$ by means of the Gaussian
approximation. To regularize the divergency of the integral over
$a_\mu$ caused by the gauge symmetry, it is suitable to fix the
background gauge
\begin{equation}
D_\mu( ^\phi\bar{A})a_\mu =0\ .
\label{10}
\end{equation}
Accordingly, the associated Faddeev-Popov determinant is to be included
into the path integral measure. As a result, we obtain an effective
scalar field theory for the dynamics in the dominant sector $[\bar{A}]$.

Some remarks are in order. When performing an Abelian projection,
some monopole distributions may multiply occur, meaning that some
gauge non-equivalent potentials $A_\mu$ may lead to the same monopole
distribution $q_\chi$. A monopole configuration should be taken with the
weight equal to its multiplicity in the Wilson loop average over
monopoles. In the effective scalar field theory constructed above,
fluctuations of the monopole distribution $q_\phi(x)$ are described
by angular variables $\Omega_\phi$ and zeros of $\phi$, while fluctuations
of the eigenvalues of $\phi$ with a fixed distribution of zeros are
associated with fluctuations of the multiplicity for a given $q_\phi$.

\subsection*{\bf 6. Conclusions}

Assuming that the monopole dominance discovered
in the lattice QCD survives the continuum limit, we have constructed
an effective field theory for the monopole dynamics which should
describe the behavior of QCD at large distances (confinement). We
have also proposed a gauge (or projection) invariant formulation of
the monopole dominance. The effective theory is therefore gauge
invariant and respects the Lorentz symmetry.

It is worth mentioning
that our approach does not suffer from the Gribov ambiguity in contrast
to, for example, the maximal Abelian projection where the Gribov problem
has to be resolved numerically \cite{hioki} and seems hopeless to solve in
the continuum limit. Note that the Gribov problem is irrelevant in the
background gauge (\ref{10}) because the monopole dominance justifies
the perturbation expansion over $a_\mu$.

The auxiliary scalar field $\phi$ can also be viewed as a {\it field
collective coordinate} para\-met\-rizing the dominant configurations
in QCD at large distances, whereas the trick of inserting the
identity (\ref{6}) into the gauge field path integral is nothing
but a way to obtain a right gauge and Lorentz invariant measure
for the collective coordinate.

\subsection*{Acknowledgments}
The author is thankful to M. Asorey, A. Hart,
H. Markum, J. Mourao, M.I. Polikarpov and F.G. Scholtz for fruitful
discussions and their interest in this work.

\end{document}